# Morphology in the era of

The study of galaxies has changed dramatically over the past few decades with the advent of large-scale astronomical surveys. These large collaborative efforts have made available high-quality imaging and spectroscopy of hundreds of thousands of systems, providing a body of observations which has significantly enhanced our understanding not only of cosmology and large-scale structure in the universe but also of the astrophysics of galaxy formation and evolution. Throughout these changes, one thing that has remained constant is the role of galaxy morphology as a clue to understanding galaxies. But obtaining morphologies for large numbers of galaxies is challenging; this topic, "Morphology in the era of large surveys", was the subject of a recent discussion meeting at the Royal Astronomical Society.

Morphology has been recognized as an important tool in categorizing and characterizing galaxies for more than a century, dating back to the early work of Hubble and others. As **Roger Davies** (University of Oxford) reminded us in an invited talk to open the meeting, the shape of a galaxy is essentially a measure of its orbital structure as determined by its gravitational potential. Gas, dust and stars pile up at orbital resonances, and the art of analysing morphology becomes one of interpreting what can be seen or measured in images as a frozen snapshot of the dynamics of the different components.

For many astronomers, morphological classification involves placing objects on some variant of the Hubble tuning-fork diagram (figure 1). Alternative classification schemes have been developed, but most share basic features with Hubble's seminal work; most often the tuning fork version enhanced by Sandage and collaborators starting in the 1960s is used. The subtly different scheme, developed by de Vaucouleurs, is also commonly applied. This emphasizes the continuity between classifications by introducing transitional morphologies; it has not only SA (unbarred spiral) and SB (barred spiral) systems, but also SAB systems with intermediate bar strengths, for example.

The fundamental division in galaxy morphology in any classification scheme is between late-type (spiral) galaxies and early-type (elliptical) galaxies. Early-type galaxies, those with a high degree of central concentration and no structure, might appear simple, but Prof. Davies reminded the conference that their smooth light distribution can hide a wide variety of orbital structures. For a smooth object where neither the intrinsic shape nor the inclination is known, the visual appearance may be completely uninformative;

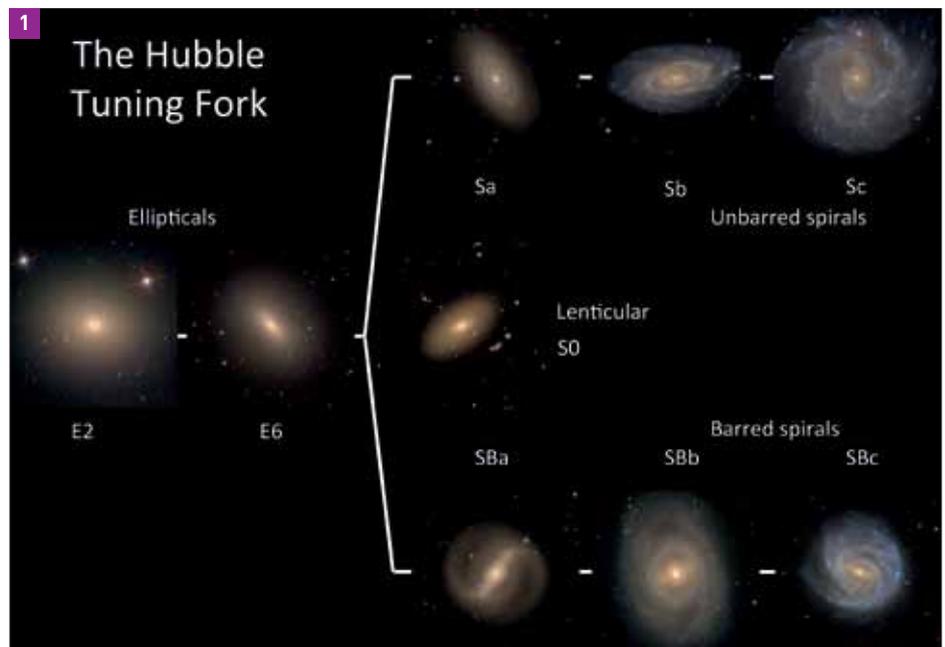

**1:** The Hubble tuning-fork diagram is the basis for most galaxy classification schemes. (Sloan Digital Sky Survey [SDSS] and K Masters, Univ. of Portsmouth) for example, a round smooth galaxy in an image could be a spherical E0, or a face-on featureless disc (S0). Structure within these systems has been studied since at least the late 1970s (Carter 1978) when systems were divided according to their isophotal shapes into "boxy" or "disky" systems; more recent work (e.g. Kormendy and Bender 2012) has led to the suggestion that the Hubble diagram's low-luminosity tail may need alteration. Complicating the picture further, the resolved spectroscopy survey ATLAS3D measured the stellar kinematics of the early-type population to show that many galaxies morphologically classed as E in fact rotate like discs. As an example, Prof. Davies showed us that NGC 3379 and NGC 4636 look almost identical in their optical images, but the kinematics of the rounder looking NGC 3379 (figure 2) reveal a rotating disc, while the slightly elongated NGC 4636 hides no rotation.

Late-type disc-dominated galaxies have a more obvious three-dimensional shape, and corrections can be made for viewing angle to reveal the face-on structures. The internal structures in disc galaxies can then be used to great effect to reveal the details of their secular evolution (a term introduced by Kormendy in the 1970s, meaning evolution that is slow compared to the dynamical timescale). The second invited talk of the morning, by **Lia Athanassoula** (Laboratoire d'Astrophysique de Marseille) concentrated on the theoretical understanding of bars in disc galaxies, and in particular on the links between theory and observation. Prof. Athanassoula began by emphasizing that when it comes to understanding disc galaxy dynamics, details matter: the presence of a bar, for example, rather than being like "frosting on the cake", might drive the evolution of a system, and the presence of spirals and rings (even if they really are just "frosting") allows the direct testing of theoretical models of stellar orbits. Prof. Athanassoula showed how theoretical models that assume a galaxy potential and use manifold methods have had remarkable success at reproducing not only the broad features of bars in disc galaxies, but even the details of the bars' internal structure, for example, if it has a rectangular shape, or is more oval.

More recent simulations, such as those in Athanassoula et al. (2013) are able to take into account the effect of dynamical dark matter halos and the cosmic evolution of the gas supply. Having done so, models with only a few parameters can reproduce even the subtle features of objects such as NGC 936 (figure 4); the apparent bulge at the centre of this system could be part of the bar (another reminder of the difficulty of distinguishing subtle morphological features). This work also makes predictions about population statistics; because it shows that bars take longer to form in the presence of classical bulges and/or in gas-rich galaxies, we should see the resulting trends of the bar fraction in the disc galaxy population. Indeed, several recent works have shown bars are more

# large surveys

**Meeting report** Chris Lintott and Karen Masters review progress in understanding galaxy morphology, as discussed at this RAS meeting.

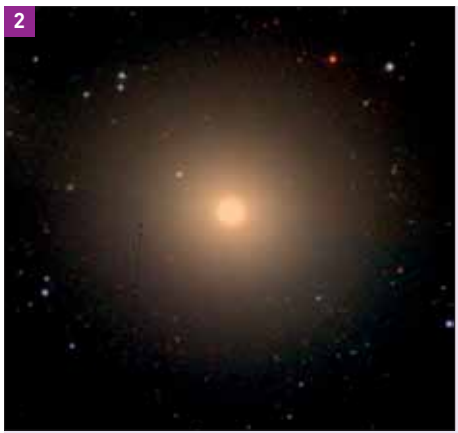

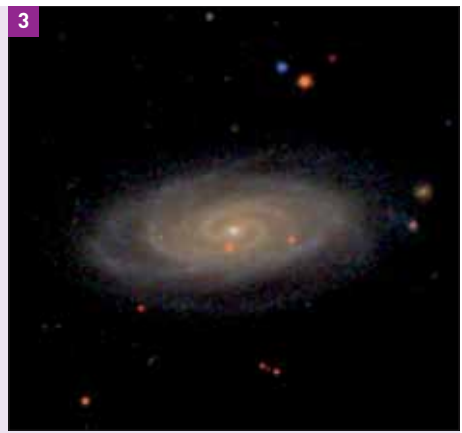

**2:** In the optical, NGC 3379 looks like a typical elliptical galaxy; ATLAS 3D revealed the smooth structure hides a rotating disc.

**3:** IC 756 is a remarkable galaxy. Despite having no visible bulge, it hosts an actively growing supermassive black hole.

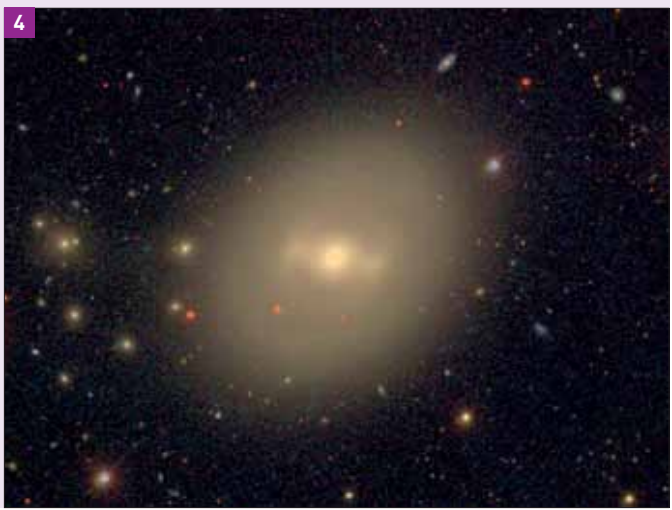

**4:** What looks like a bulge at the centre of NGC 936 might actually be part of the bar.
(All SDSS)

common in more massive, redder and less star-forming disc galaxies, as we shall see below.

## Observing morphologies

Several of the effects mentioned above have been noticed recently thanks to the large morphological data sets that are now available. Indeed, much of the renewed interest in morphology may be due to the availability of such resources. Among the largest such datasets are those produced by the Galaxy Zoo citizen science project (http://www.galaxyzoo.org). In the first contributed talk of the day, **Kyle Willett** (University of Minnesota) presented results from Galaxy Zoo 2, which provides detailed morphological classifications of more than 300 000 galaxies drawn from the Sloan Digital Sky Survey (Willett *et al.* 2013). Defining a successful classification as a measurement of morphology, Willett noted that automatic classifiers too often use colour as a proxy for morphology. If the central reason to be interested in morphology is to reveal the dynamical history of the system in question, the colour – which measures primarily only the recent star formation history – even if it correlates very well with morphology, is a distraction. Willett described techniques used to collate the ~40 votes per question, correct for observational biases, and create "classification likelihoods" from the Galaxy Zoo project. The classifications, which are now available via the Galaxy Zoo site (http://data.galaxyzoo.org) and as part of the Tenth Data Release of SDSS, provide interesting insight on what experts really mean when classifying galaxies. For example, GZ2 classifications show no strong trend between the pitch angle of the spiral arms and the T-type of galaxies as recorded by experts such as Nair and Abraham (2010). In other words, classical spiral Hubble types as recently used in the astronomical literature have evolved to mean measurements almost exclusively of bulge dominance, rather than also accounting for the tightness of the winding of the spiral arms.

In the last invited talk of the morning, **Steven Bamford** (University of Nottingham) looked beyond Galaxy Zoo to the issue of automated methods for measuring morphology. Laying out the requirements for a perfect measure, he demanded that it should be objective and efficient, yet sensitive to subtle features. It should also be understandable: black boxes are often unhelpful when physical insight is sought. These criteria could be used to assess successive generations of attempts, from basic single-parameter metrics such as colour, concentration of Sersic index, through to higher-order attempts such as CAS (concentration, asymmetry and clumpiness; the main features it assesses) or Gini-$M_{20}$ which also make use of multiple parameters.

Perhaps the most substantial progress towards automated morphology has been attempts to deploy neural networks for galaxy classification. Lahav *et al.* (1995) were already able to use such methods to match the performance of a set of expert classifiers, although at the cost of using inputs that included manual measurements, and requiring a training set that included most of the subjects to be classified. Ball *et al.* (2004) applied neural networks to SDSS galaxies for the first time, but included colour, magnitude and size in the input parameters. These variables correlate well with morphology, but are not themselves morphological, and when included they dominate the classification. A hybrid approach taken by Banerji *et al.* (2010) used Galaxy Zoo morphologies to train and validate automated classifiers. They also attempted to use only morphological inputs, with reasonable success. So far, surprisingly, sophisticated quantities expected to correlate well with morphology, such as "texture", provide much less discriminatory power than simple measures of the concentration of the light profile.

For most galaxies, automated approaches are already as accurate as human classifiers for basic morphology. The main remaining difficulties are identifying objects that do not follow the typical trends – such as spirals with dominant bulges, or discs without spiral arms – and dealing with galaxies with uncertain or intermediate classifications. These are often the most important populations for improving our physical understanding of galaxies. But should we be aiming to do better than reproducing human classifications? It was posited that an ideal automated morphological classifier should be trained using as much information as possible, including high-quality imaging and kinematic data, then perform its classifications using only uncalibrated images, to prevent it from relying on non-morphological information. This would make an interesting challenge for the machine-learning community.

Finally, this talk highlighted several automated approaches that attempt to go beyond classification and measure the properties of specific morphological features such as bars, spiral arms, bulges and discs. In particular, the advantages of harnessing the availability of the latest large, multi-wavelength datasets were demonstrated. A lot is undoubtedly still to be done to produce measures of morphology capable of making the most of the huge quantity of data provided by modern surveys, but the work is underway, and much science is flowing from the classifications we already have.

### Science with morphology

With its involvement of hundreds of thousands of volunteers, the Galaxy Zoo project has been effective public outreach, but the project's long-term significance also depends on the science it enables, a point highlighted by the fact that most volunteers say they spend time on the project to help science (Raddick *et al.* 2010). **Karen Masters** (University of Portsmouth) began the presentation of science results from Galaxy Zoo, returning to the theme of bars. She talked about results which show that while perhaps as many as 75–80% of red spirals have strong bars of the type found by the survey, the same is not true for the blue population, an important effect previously obscured by the practice of selecting disc galaxies out of large galaxy samples by colour.

This result raises the interesting possibility that bars might be contributing to the cessation of star formation in red disc galaxies (or the bar is a side effect of processes that turn the disc galaxy red).

To look directly at the connection between the bar and gas in a galaxy, Masters used data from ALFALFA, a blind HI survey conducted with the Arecibo radio telescope which now covers approximately half of the SDSS Legacy area. The net result is that more bars are seen in systems which have less HI, and that at fixed galaxy mass it is the gas fraction that dominates the odds of a particular disc having a bar. In reverse, for a given gas fraction, barred systems are observed to be more likely to be red than their non-barred counterparts (Masters *et al.* 2012).

Taking advantage of the large sample sizes from Galaxy Zoo and ALFALFA, Masters identified some very rare, gas-rich, strongly barred discs, and has been awarded Jansky Very Large Array time to obtain resolved HI images of some. Curiously, these have extreme gas fractions but a wide variety of optical colours.

Observing bars at higher redshift was the topic of the next talk by **Thomas Melvin** (University of Portsmouth), using data from the third phase of the Galaxy Zoo project, which provides morphological classifications for galaxies drawn from the HST COSMOS field. This is of especial interest because bars are believed to form in dynamically stable galaxies, and so an increasing bar fraction with time (or decreasing with redshift) is a tracer for a maturing disc galaxy population. Although care needs to be taken to avoid observational biases, this effect is exactly what is seen in the new Galaxy Zoo data.

Turning from the evolution of a particular morphological feature to the trajectory taken by systems throughout their history, **Kevin Schawinski** (ETH Zurich) discussed the morphologies of galaxies crossing the green valley. The colour–magnitude (or, equivalently, colour–mass) diagram has become an important tool for understanding galaxy evolution, and the bulk of the local galaxy population lies in regions of the space called the red sequence and the blue cloud; the green valley is the relatively unpopulated region between these. Schawinski used galaxies with well-determined morphologies from Galaxy Zoo 1 to show how, in the colour–magnitude diagram, the green valley late and early types appear only as the tails of their distribution. Schawinski shows that, however, the green valley late-types are still a tail in the distribution of star formation versus mass, but green valley early-types are displaced from the main population.

Schawinski's recently submitted work uses UV/optical colours, corrected for dust, in order to constrain the star formation history of the systems. Models that assume a constant star formation rate and appropriate quenching provide a simple set of toy models with which to test these ideas, revealing the need for a sudden quenching of star formation to move early-type systems onto the red sequence. If such rapid quenching indeed proves to be necessary, then it may be evidence for such processes as merger-induced AGN feedback. By contrast, there is no evidence for such a sudden cut-off in late-types, which might be expected to move only slowly onto the green valley itself. Schawinski proposes this is consistent with models in which the gas supply is shut off, but no star formation quenching happens.

### Do mergers matter?

These observations support the now standard model in which early-type galaxies are merger products, with a rapid shut-off of star formation as a result of AGN feedback triggered by gas inflow from the interaction. The contribution of mergers to the assembly of galaxies was the subject of much debate during the meeting, coming up in several contributed talks. Speaker **Sugata Kaviraj** (University of Hertfordshire) presented attempts to test the traditional model, which sees major mergers driving the transformation of galaxies. Using recent WFC3 data to probe the peak of star formation density at a redshift of about 2, Kaviraj *et al.* (2013) find, perhaps surprisingly, that the star formation rate of major mergers are not significantly different from the background population. Only 15% of star formation at this epoch is in systems which show morphological signatures of a major merger. As well as star formation in disc galaxies, we see a substantial number of blue spheroids devoid of tidal features but which appear to have rapid star formation timescales. These, it seems, are systems in the main phase of their assembly: spheroids forming directly from collapse of clumpy discs. Surveys such as CANDELS provide the promise of very soon going beyond these relatively small-number statistics and surveying the bulk of the population of normal star forming galaxies at high redshift.

If we are to understand the contribution of mergers to galaxy assembly, we ideally need a control sample of galaxies which have been untouched by major interactions. Such a sample was presented by **Brooke Simmons** (University of Oxford), who showed a sample of the most bulge-free galaxies selected from Galaxy Zoo. Simulations show that any significant interaction (perhaps even down to a mass ratio of 1:10) would inevitably kick stars up into a bulge, and so bulgeless galaxies, with less than 1% of their light in either a bulge or pseudobulge, should be guaranteed merger-free.

Amazingly, Simmons presented a sample of 13 such bulgeless galaxies with actively growing black holes (optical AGN; e.g. figure 3). Perhaps not surprisingly, these bulgeless galaxies lie off the standard bulge mass–black hole mass relation, but are found on the line for total stellar mass–black hole mass relation. In other words, these systems may be telling us that the growth of a galaxy is controlled by the dark matter halo potential, rather than its merger history. These results (Simmons *et al.* 2013) are based on a small sample of the most bulgeless systems with AGN found in Galaxy Zoo; the next step is to establish how rare such merger-free systems are, along with detailed follow-up at Gemini and WIYN.

If these results indeed show that the growth of the central black hole in most galaxies is driven not by merger-triggered accretion but rather by secular processes, this presents a challenge to modellers. This problem was the subject of a talk by **Victor Debattista** (University of Central Lancashire), who presented a set of models that show the disc and the bulge growing together over time: as the disc grows, the bulge becomes compressed and its velocity dispersion increases. (Debattista *et al.* 2013)

Taking into account these effects, the models predict that a black hole should grow at about half the rate of the disc, a number that is roughly in agreement with studies such as that carried out by Gadotti and Kauffman (2009). However, the predicted M-sigma relation (e.g. Gültekin *et al.* 2009) does not fit observations. Despite the lack of mergers, the black hole must still be growing in a self-regulating way, with gas falling directly into the centre.

The discussion stimulated by the sample of unusual (bulgeless) galaxies highlighted the need to classify large data sets; just 13 bulgeless galaxies with AGN were ultimately derived

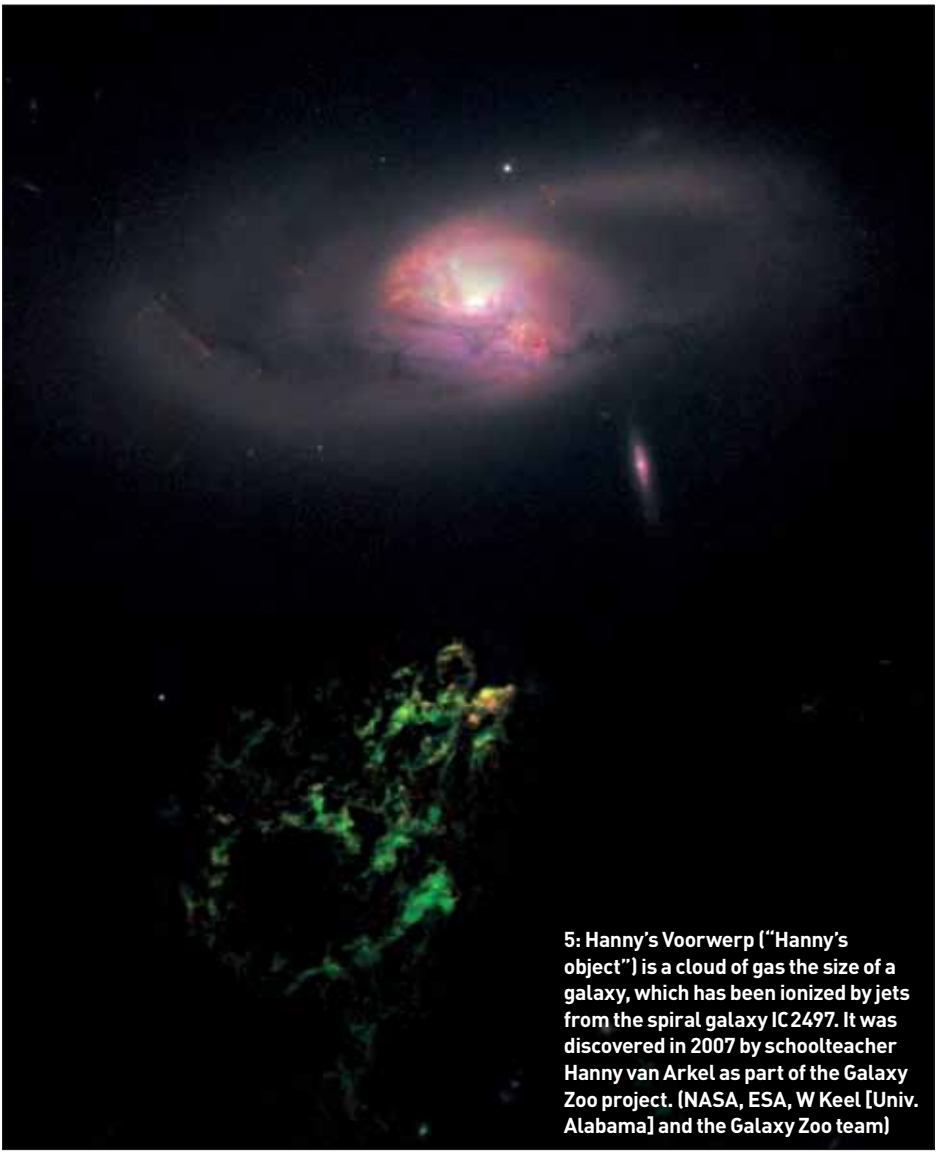

5: Hanny's Voorwerp ("Hanny's object") is a cloud of gas the size of a galaxy, which has been ionized by jets from the spiral galaxy IC 2497. It was discovered in 2007 by schoolteacher Hanny van Arkel as part of the Galaxy Zoo project. (NASA, ESA, W Keel [Univ. Alabama] and the Galaxy Zoo team) from the Galaxy Zoo sample of nearly 900 000 systems. One of the biggest successes of the Galaxy Zoo project has been the serendipitous discovery of truly unusual objects, an area highlighted by **Chris Lintott** (University of Oxford) who spoke about what we can learn from light echoes of AGN. The first of these extremely rare objects was found by Galaxy Zoo volunteers and dubbed Hanny's Voorwerp (figure 5; Lintott *et al.* 2009). This object has been shown to be a galaxy-scale gas cloud highly ionized by jets associated with AGN activity in IC 2497, a neighbouring infrared–luminous disturbed spiral galaxy. However, X-ray and infrared observations have established that IC 2497's current AGN luminosity is insufficient to account for the Voorwerp's appearance. This suggests that a drop of at least an order of magnitude in AGN luminosity has taken place within the past few hundred thousand years in IC 2497. Inspired by this discovery, Galaxy Zoo volunteers had found more examples of AGN-ionized gas clouds (Keel *et al.* 2012), which have been the subject of an extensive HST programme. Of the 19 studied in detail, it is exciting to note that more than a third show signs of rapid shutdown of the AGN, perhaps representing a shift from radio to kinetic modes of accretion, and a laboratory to study the most recently quenched AGN in the local universe.

## Morphology at high redshift

The meeting closed with an invited talk by **Jennifer Lotz** (Space Telescope Science Institute) on galaxy morphology in large samples at high redshift. Lotz reminded the meeting that morphology at high redshift, particularly identifying merging systems, is a tool to understand how hierarchical assembly works on small scales to grow galaxies (see Dekel *et al.* 2009). Following a series of talks about the possible relative unimportance of merging, Lotz reminded us that at least they can be observed and counted at high redshift; while gas accretion may play a significant role in galaxy evolution, it is impossible to observe directly. Identifying mergers consistently, however, is also not easy: confusion with close non-merging pairs is always a problem. Lotz's use of the Gini-$M_{20}$ method (which measures uniformity and concentration in an image) provides an objective measure of merger likelihood. Using this method, the fraction of mergers in the galaxy population doesn't seem to evolve out to a redshift of 1.2.

At $z < 1.5$, spheroids can be formed by mergers but, as noted by previous speakers, they cannot be the major drivers of either star formation or AGN activity. Determining merger fractions above this redshift is difficult, and simulations of the sort used to calibrate the more local data will undoubtedly be important. Data from both CANDELS and from Herschel show that there is lots of merging, but also that low-luminosity AGNs are ubiquitous by $z = 2$. To reconcile all of this information requires work, but a picture is emerging of two pathways to quiescent spheroids as seen in the local universe: one merger driven, and one through isolated collapse.

There is undoubtedly much more to come from this work, and morphology – a subject introduced at the dawn of galactic astrophysics – is still a critical component of our attempts to understand the galaxy population we see around us. A stimulating and lively meeting brought together observers, classifiers, simulators, modellers and theorists to discuss what remains to be done. Perhaps the most important conclusion is that these groups now have a shared understanding of what morphology means. We should expect to hear much more about mergers and bars, about bulgeless galaxies and spheroidals for the foreseeable future. ●

**References**
**Athanassoula E** *et al.* 2013 *Mon. Not. R. Astron. Soc.* **429** 1949.
**Ball N M** *et al.* 2004 *Mon. Not. R. Astron. Soc.* **348** 1038.
**Banerji M** *et al.* 2010 *Mon. Not. R. Astron. Soc.* **406** 432.
**Carter D** 1978 *Mon. Not. R. Astron. Soc.* **182** 797.
**Debattista V P** *et al.* 2013 *Ap. J.* **765** 23.
**Dekel A** *et al.* 2009 *Ap. J.* **703** 785.
**Gadotti D A and Kauffman G** 2009 *Mon. Not. R. Astron. Soc.* **399** 621.
**Gültekin K** *et al.* 2009 *Ap. J.* **706** 404.
**Kaviraj S** *et al.* 2013 *Mon. Not. R. Astron. Soc.* **429** L40.
**Keel W C** *et al.* 2012 *Mon. Not. R. Astron. Soc.* **420** 878.
**Kormendy J** and **Bender R** 2012 *Ap. J. Supp.* **198** 2.
**Lahav O** *et al.* 1995 *Science* **267** 859.
**Lintott C** *et al.* 2009 *Mon. Not. R. Astron. Soc.* **399** 129.
**Masters K** *et al.* 2012 *Mon. Not. R. Astron. Soc.* **424** 2180.
**Nair P B and Abraham R G** 2010 *Ap. J. Supp.* **186** 427.
**Raddick M J** *et al.* 2010 *Astronomy Education Review* **9** 010103.
**Schawinski K** *et al. Mon. Not. R. Astron. Soc.* submitted.
**Simmons B D** *et al.* 2013 *Mon. Not. R. Astron. Soc.* **429** 2199.
**Willett K** *et al.* 2013 *Mon. Not. R. Astron. Soc.* in press.